\documentclass[figures,preprint]{epl}   
\usepackage{amsfonts}
\usepackage{amssymb}
\usepackage{amsbsy}
\usepackage{graphics}
\usepackage{psfig}

\input epsf.sty

\newcommand{\BEQ}{\begin{equation}}     
\newcommand{\BEA}{\begin{eqnarray}}
\newcommand{\EEQ}{\end{equation}}       
\newcommand{\EEA}{\end{eqnarray}}
\newcommand{\eps}{\varepsilon}          
\newcommand{\D}{{\rm d}}                
\newcommand{\II}{{\rm i}}               
\newcommand{\wit}[1]{\widetilde{#1}}    

\renewcommand{\vec}[1]{\boldsymbol{#1}} 

\title{Two-time autocorrelation function in phase-ordering
kinetics from local scale-invariance}
\shorttitle{Autocorrelation function in phase-ordering kinetics~}

\author{Malte Henkel\inst{1}, Alan Picone\inst{1}, Michel Pleimling\inst{2}}

\institute{
\inst{1}Laboratoire de Physique des Mat\'eriaux (CNRS UMR 7556), 
Universit\'e Henri Poincar\'e \\ Nancy I, B.P. 239,
F -- 54506 Vand{\oe}uvre l\`es Nancy Cedex, France\\
\inst{2} Institut f\"ur Theoretische Physik I, 
Universit\"at Erlangen-N\"urnberg,\\
D -- 91058 Erlangen, Germany}
\pacs{05.70.Ln}{Nonequilibrium and irreversible thermodynamics}
\pacs{64.60.Ht}{Dynamic critical phenomena}
\pacs{11.25.Hf}{Conformal field-theory}

\begin{document}
\maketitle
\begin{abstract}
The time-dependent scaling of the two-time autocorrelation function of 
spin systems without disorder undergoing 
phase-ordering kinetics is considered. Its form
is shown to be determined by an extension of dynamical scaling to a local
scale-invariance which turns out to be a new version of conformal invariance. 
The predicted autocorrelator is in agreement with Monte-Carlo data on the
autocorrelation function of the $2D$ kinetic Ising model with Glauber dynamics 
quenched to a temperature below criticality. 
\end{abstract}

Understanding the kinetics of phase-ordering after a rapid quench from an 
initial disordered state into the ordered phase has since a long time 
posed a continuing challenge 
(see \cite{Bray94,Godr02,Cugl02,Henk04} for reviews). 
A key insight has been the 
observation that many of the apparently erratic and history-dependent 
properties of such systems can be organized in terms of
a simple scaling picture \cite{Stru78}. This means
that there is a single time-dependent length-scale $L(t)$ which is 
identified with the typical linear size of ordered clusters. It turns
out that the ageing behaviour is more fully revealed in 
observables such as the two-time
autocorrelation function $C(t,s)$ or the two-time linear 
autoresponse function $R(t,s)$ defined as
\BEQ \label{gl:CR}
C(t,s) := \langle \phi(t) \phi(s) \rangle \;\; , \;\; 
R(t,s) := \left.\frac{\delta \langle \phi(t)\rangle}{\delta h(s)}\right|_{h=0}
\EEQ
where $\phi(t)$ denotes the time-dependent order-parameter, $h(s)$ is the
time-dependent conjugate magnetic field, $t$ is referred to as {\em observation 
time} and $s$ as {\em waiting time}. One says that the system undergoes
{\em ageing} if $C$ or $R$ depend on both $t$ and $s$ and not merely on the
difference $\tau=t-s$. These two-time functions are expected to show dynamical
scaling in the ageing regime $t,s\gg t_{\rm micro}$ and $t-s\gg t_{\rm micro}$,
where $t_{\rm micro}$ is some microscopic time scale. Then 
\BEQ \label{gl:SkalCR}
C(t,s) = M_{\rm eq}^2 f_C(t/s) \;\; , \;\; R(t,s) = s^{-1-a} f_R(t/s)
\EEQ
such that the scaling functions $f_{C,R}(y)$ satisfy the following asymptotic
behaviour
\BEQ \label{gl:lambdaCR}
f_C(y) \sim y^{-\lambda_C/z} \;\; , \;\; 
f_R(y) \sim y^{-\lambda_R/z}
\EEQ
as $y\to\infty$ and where $\lambda_C$ and $\lambda_R$, respectively, are known
as the autocorrelation \cite{Fish88,Huse89} and autoresponse exponents
\cite{Pico02} and $z$ is the dynamical exponent, defined through 
$L(t)\sim t^{1/z}$. Throughout, we consider simple ferromagnets without 
disorder and with a non-conserved order-parameter. 
Then $z=2$ is known \cite{Rute95b}. For spin systems with short-ranged
equilibrium correlators (e.g. the $d>1$ Glauber-Ising model) it has been 
checked in detail that $a=1/z=1/2$ \cite{Henk02a,Henk03e}, 
but the closed-form and {\it ad hoc} OJK approximation gives 
$a=(d-1)/2$ \cite{Bert99,Maze98}. 
The exponents $\lambda_{C,R}$ are 
independent of the equilibrium exponents and 
of $z$ \cite{Jans89,Godr02,Cugl02}. Although the equality 
$\lambda_C=\lambda_R$ had been taken for granted 
(reconfirmed in a recent second-order perturbative analysis
of the time-dependent Ginzburg-Landau equation \cite{Maze03}), 
counterexamples exist for 
long-ranged initial correlations in ageing ferromagnets \cite{Pico02} 
and in the random-phase sine-Gordon model \cite{Sche03}. 
For {\em short-ranged} initial correlations,
dynamical scaling together with Galilei-invariance at temperature 
$T=0$ are sufficient for $\lambda_C=\lambda_R$ \cite{Pico04}.

We are interested in the form of the scaling functions $f_{C,R}(y)$.
Indeed, it is known that for any given value of $z$ there exist
infinitesimal local scale-transformations 
$t\mapsto (1+\eps)^z t$, $\vec{r} \mapsto (1+\eps)\vec{r}$ 
with an infinitesimal $\eps=\eps(t,\vec{r})$ which  
may depend on both time and space \cite{Henk02}.
Furthermore, the {\em local} scale-transformations
so constructed act as dynamical symmetries of certain linear field equations
which might be viewed as some effective renormalized equation of motion. 
{}From the assumption that the response functions of the
theory transform covariantly under local scale-transformations, the exact
form of the scaling function $f_R(y)$ is found \cite{Henk02,Henk01,Pico04}
\BEQ \label{gl:fR}
f_R(y) = r_0 y^{1+a'-\lambda_R/z} \left( y-1 \right)^{-1-a'}
\EEQ
where $a'$ is a new exponent \cite{Pico04} and $r_0$ is a normalization 
constant. Eq.~(\ref{gl:fR}) with $a=a'$ is recovered 
in many spin systems quenched to a temperature $T\leq T_c$ and whose
dynamics is described by a master 
equation \cite{Henk02,Henk01,Henk03b,Henk04,Pico04,Plei04}, but in models
such as the $1D$ Glauber-Ising model at $T=0$ \cite{Pico04} 
or the OJK approximation of phase-ordering \cite{Bert99,Maze98,Maze03} 
eq.~(\ref{gl:fR}) holds but with $a\neq a'$. 
If a phase-ordering system is also Galilei-invariant at $T=0$, then $f_R(y)$ is
independent of both the thermal and the initial noises \cite{Pico04}. 

While the scaling form of the autoresponse function thus seems to be
understood, the problem of finding the scaling function $f_C(y)$ of the
autocorrelation function appears to be considerably more difficult. A 
by now classical attempt recognizes that for $T<T_c$, temperature should
be irrelevant \cite{Bray94} and hence sets $T=0$. 
Building on the Ohta-Jasnow-Kawasaki 
approximation (see \cite{Bray94}) in the kinetic O($n$)-model one 
introduces an auxiliary field for which a gaussian closure procedure is
assumed. This leads to \cite{Bray91a,Bray92,Roja99}
\BEQ \label{gl:BPT}
f_{C,{\rm BPT}}(y) = \frac{n}{2\pi} 
\left[ B\left(\frac{1}{2},\frac{n+1}{2}\right)\right]^2 
\left(\frac{4y}{(y+1)^2}\right)^{d/4}
{_2F_1}\left(\frac{1}{2},\frac{1}{2};\frac{n+2}{2};
\left(\frac{4y}{(y+1)^2}\right)^{d/2}\right) 
\EEQ
where $B$ is Euler's beta function and ${_2F_1}$ a hypergeometric function.
However, this closed form implies $\lambda_C=d/2$ which only holds in certain 
limiting cases (for example, $\lambda_C=d/2+\alpha n^{-1}$, to leading order 
in $n$, in the O($n$)-model and with a known value of
$\alpha>0$ \cite{Bray91}. See also \cite{Brow97}).  

Here we investigate to what extent $f_C(y)$ may be determined from a
local scale-invariance (LSI). We concentrate 
on phase-ordering where $T<T_c$ and thus
$z=2$ \cite{Rute95b}. The group of local scale-transformations is then the
Schr\"odinger group \cite{Nied72,Hage72} which for example arises as the
maximal kinematic group of the free Schr\"odinger (or diffusion) equation. 
In particular, the Schr\"odinger group contains dilatations with $z=2$ and
Galilei-transformations. For local theories, 
there is a Ward identity such that these two symmetries imply full 
Schr\"odinger-invariance \cite{Henk03}. However, Galilei-invariance is
incompatible with thermal or initial noises. 

We consider a coarse-grained order-parameter $\phi(t,\vec{r})$
satisfying a Langevin equation \cite{Hohe77,Bray94} 
\BEQ \label{gl:Lang}
\frac{\partial \phi(t,\vec{r})}{\partial t}=
-D\frac{\delta {\cal H}}{\delta \phi}-Dv(t)\phi(t,\vec{r})+\eta(t,\vec{r})
\EEQ
where $\cal H$ is the classical Hamiltonian, and $D$ stands for the diffusion 
constant. Zero-mean thermal noise is characterized by its variance 
$\langle\eta(t,\vec{r})\eta(s,\vec{r}')\rangle=
2DT\,\delta(t-s)\,\delta(\vec{r}-\vec{r}')$ 
where $T$ is the bath temperature. The initial conditions are specified
in terms of $a(\vec{r}-\vec{r}') := 
\left\langle \phi(0,\vec{r}) \phi(0,\vec{r}')  \right\rangle$ 
and where we already anticipated spatial translation invariance, hence
$a(\vec{r})=a(-\vec{r})$. The potential $v=v(t)$ acts as a
Lagrange multiplier. For $z=2$ it is easy to see that if 
\BEQ
k(t) := \exp\left( -D\int_0^t \!\D u\: v(u) \right) \sim t^{\digamma} \;\; , 
\;\; \digamma = 1+a - \frac{\lambda_R}{2}
\EEQ
and if $a=a'$ then eq.~(\ref{gl:fR}) is reproduced from 
Schr\"odinger-invariance \cite{Pico04}. The Langevin equation
(\ref{gl:Lang}) may be turned into a field-theory using the
Martin-Siggia-Rose formalism. Provided that field-theory is 
{\em Galilei-invariant in the absence of thermal and of initial noise} 
(i.e. $T=0$ and $a(\vec{r})=0$)
then the two-time autocorrelation function can be expressed in terms
of noiseless response functions. Precise data on the form of the
space-time response function in the Glauber-Ising model in $2D$ and $3D$ 
provide strong direct evidence in favour of its 
Galilei-invariance \cite{Henk03b}. We concentrate on the
case of a fully disordered initial state with 
$a(\vec{R})=2a_0\delta(\vec{R})$ where $a_0$ is a normalization constant. 
Then it is shown in \cite{Pico04} that
\BEA
C(t,s) &=& a_0\int \!\D\vec{R}\: R_{0}^{(3)}(t,s,0;\vec{R})
+DT\int \!\D u\,\D\vec{R}\: R_{0}^{(3)}(t,s,u;\vec{R})
\label{glC1}
\\
R_{0}^{(3)}(t,s,u;\vec{r}) &:=&
\left\langle\phi(t;\vec{y})
\phi(s;\vec{y}){\wit \phi}(u;\vec{r}+\vec{y})^2\right\rangle_0
\:=\: \frac{k(t)k(s)}{k^{2}(u)} {\cal R}_{0}^{(3)}(t,s,u;\vec{r})
\label{glC2}
\EEA
where the index 0 refers to the noiseless part of the field-theory. 
Here, the field $\phi$ has the scaling dimension $x=1+a$ and 
$\wit{\phi}^{\,2}$ is a composite field with scaling dimension 
$2\wit{x}_2$ (only for free fields $\wit{x}_2 =x$).
The well-known three-point response function ${\cal R}_0^{(3)}$ for 
$v(t)=0$ is given by Schr\"odinger-invariance \cite{Henk94}
\BEA
{\cal R}_{0}^{(3)}(t,s,u;\vec{r}) &=& {\cal R}_{0}^{(3)}(t,s,u)
\exp\left[-\frac{{\cal M}}{2}\frac{t+s-2u}{(s-u)(t-u)}{\vec{r}}^{2} \right]
\Psi\left(\frac{t-s}{2(t-u)(s-u)}{\vec{r}}^{2}\right)
\nonumber \\
{\cal R}_{0}^{(3)}(t,s,u) &=& \Theta(t-u)\Theta(s-u)\,
\left(t-u\right)^{-{\wit x}_2}
\left(s-u\right)^{-{\wit x}_2}\left(t-s\right)^{-x+{\wit x}_2}
\label{glC3}
\EEA
where $\Psi=\Psi(\rho)$ is an arbitrary scaling function 
and ${\cal M}=1/(2D)$ is a 
non-universal constant. Eqs.~(\ref{glC1},\ref{glC2},\ref{glC3}) are the 
foundation of our analysis of the autocorrelation function.

Comparing (\ref{glC1},\ref{glC2},\ref{glC3}) with the scaling form
(\ref{gl:SkalCR},\ref{gl:lambdaCR}), we have 
$\wit{x}_2-x=d/2-\lambda_C$ and \cite{Pico04}
\BEQ \label{gl:CPhi}
C(t,s) = {a_0} y^{\lambda_{C}/2}\left(y-1\right)^{-\lambda_C}
\Phi\left(\frac{y+1}{y-1}\right) \;\; , \;\;
\Phi(w) := \int_{\mathbb{R}^d} 
\!\D\vec{R}\:\exp\left[-\frac{{\cal M}w}{2}\,{\vec R}^2\right]
\Psi\left({\vec R}^{2}\right)
\EEQ 
where $y=t/s$. The second term in (\ref{glC1}) merely gives a finite-time 
correction and may be dropped, in agreement with $T$ being irrelevant for
$T<T_c$ \cite{Bray94}. 
The form of $f_C(y)$ still depends on the unknown function $\Phi(w)$. 
A simple heuristic way to fix its form is to argue that the 
noiseless response function $R_0^{(3)}(t,s,0;\vec{r})$ which describes a 
response of the autocorrelation $C(t,s)=\langle\phi(t)\phi(s)\rangle$ should be 
non-singular at $t=s$. This leads to 
$\Phi(w) \simeq\Phi_0 w^{-\lambda_C}$ as $w\to\infty$. 
If this were valid for all $w$, we would obtain the 
following simple form \cite{Pico04}
\BEQ \label{gl:appr}
C(t,s)\approx{a_0\Phi_0}
\left({(y+1)^2}/{y}\right)^{-\lambda_C/2}
\EEQ
which at least gives the correct asymptotic behaviour as $y\to\infty$. It has
been checked that this form is exact for systems described by an underlying
free-field theory \cite{Pico04}. 

We now outline how to find the scaling function $\Phi(w)$ in (\ref{gl:CPhi})
more systematically. To achieve this by a dynamical symmetry argument, an
extension of the Schr\"odinger group used so far as dynamical group has to
be found. Indeed, when considering the dynamical symmetries of the 
free Schr\"odinger equation $(2{\cal M}\partial_t-\partial_{\vec{r}}^2)\phi=0$,
it is possible to consider also the `mass' $\cal M$ as a dynamical variable
\cite{Giul96}. Then the dynamical symmetry group extends to the
conformal group in $d+2$ dimensions \cite{Henk03,Burd73}. We postulate that,
at zero temperature, 
this {\em conformal symmetry is a dynamical symmetry of phase-ordering}. 
Now a quasiprimary field depends on three variables 
$\phi=\phi({\cal M},t,\vec{r})$.
For conformal invariance, it is sufficient that $\phi$ transforms covariantly
under the extra generators \cite{Henk03} (for simplicity we also set $d=1$)
\BEQ \label{gl:conf}
V_{-} = -\II \frac{\partial^2}{\partial{\cal M}\partial r} +
\II r \frac{\partial}{\partial t} \;\; , \;\;
N_0 = - t\frac{\partial}{\partial t} -1 
-{\cal M}\frac{\partial}{\partial {\cal M}}
\EEQ 
We look for the general form of a three-point function
${\cal R}_0^{(3)}=\langle \phi_a\phi_b\wit{\phi}_c\rangle$ 
of a theory without noise and which is conformally invariant.  
Recall that in the MSR formalism, response functions are written as
correlators of $\phi$ and $\wit{\phi}$. The response field
$\wit{\phi}$ conjugate to the field $\phi$ has the `mass' 
$\wit{\cal M}=-{\cal M}\leq 0$ \cite{Henk03} and for ${\cal R}_0^{(3)}$ 
the `mass' conservation ${\cal M}_a+{\cal M}_b+\wit{{\cal M}}_c=0$ 
holds \cite{Henk94}. The response
function is given, because of Schr\"odinger-invariance, by eq.~(\ref{glC3}).
Since now the `masses' are also considered as variables, we expect 
$\Psi=\Psi(\rho,{\cal M}_a,{\cal M}_b)$. Covariance under the
conformal group means 
$V_{-}{\cal R}_0^{(3)} = N_0{\cal R}_0^{(3)}=0$ and leads to 
\BEA
& & 
\!\!\!\!\!\!
\left( \lambda_C-a-\frac{d}{2} + {\cal M}_a\frac{\partial}{\partial {\cal M}_a} 
+\frac{\partial^2}{\partial \rho \partial{\cal M}_a}\right)\Psi = 0 \;\; , \;\;
\left( \lambda_C-a-\frac{d}{2} + {\cal M}_b\frac{\partial}{\partial {\cal M}_b} 
+\frac{\partial^2}{\partial \rho \partial{\cal M}_b}\right) \Psi = 0  
\nonumber  \\
& & ~~~~~~~~~~~~~~
\left( \lambda_C-2a-\frac{d}{2} - \rho\frac{\partial}{\partial \rho} 
+ {\cal M}_a\frac{\partial}{\partial {\cal M}_a}
+ {\cal M}_b\frac{\partial}{\partial {\cal M}_b}  \right) \Psi = 0 
\EEA
hence $\Psi = \rho^{\lambda_C-2a-d/2} K(\eta,\zeta)$ with 
$\eta=({\cal M}_a+{\cal M}_b)\rho/2$ and $\zeta=({\cal M}_a-{\cal M}_b)\rho/2$.
We need the response of the 
autocorrelator $\langle\phi_a\phi_a\rangle$, thus  
${\cal M}_a={\cal M}_b={\cal M}$, hence $\zeta=0$. Then
\BEQ
\left[ \left(2\lambda_C-d-2a\right) 
+\left(\lambda_C+1-2a-d/2\right)\partial_{\eta}
+\eta\partial_{\eta} +\eta\partial_{\eta}^2\right] K(\eta,0) =0
\EEQ
and we finally obtain the required scaling function
($\psi_{0,1}$ are arbitrary constants)
\BEA 
\Psi(\rho,{\cal M},{\cal M}) &=& \psi_0\, \rho^{\lambda_C-2a-d/2}
{_1F_1}\left(2\lambda_C-d-2a,\lambda_C+1-2a-d/2;-{\cal M} \rho\right) 
\nonumber \\
& & +\psi_1\, {\cal M}^{2a+d/2-\lambda_C}
{_1F_1}\left(\lambda_C-d/2,1+2a+d/2-\lambda_C;-{\cal M} \rho\right)
\label{gl:PsiConf} 
\EEA

\begin{table}[t]
\caption{Parameters of the autocorrelation function of the $2D$ Glauber-Ising
model. \label{tab1}}
\begin{tabular}{|c|ccc|ccc|}  \hline
$T$ & $C_{\infty}$ & $C_{\rm ref}$ & $y_{\rm ref}$ & $A$ & $B$ & $E$ \\ \hline
0.0 & 1.65(3) & 0.605 & 3.5 & -0.601 & 3.94 & 0.517 \\
1.5 & 1.69(3) & 0.48  & 5.5 & -5.41  & 18.4 & 1.24  \\ \hline
\end{tabular}
\end{table}

Before we can insert this into (\ref{gl:CPhi}), we should consider the
conditions required such that the derivation of (\ref{gl:PsiConf}) 
is valid. In particular, 
it is based on dynamical scaling and we recall the condition 
$t-s\gg t_{\rm micro}$ for its validity (similar difficulties have been
encountered before for integrated response 
functions, see \cite{Zipp00,Henk02a,Henk03e}). 
{}From (\ref{glC3}), this means that
for small arguments $\rho\to 0$ the form of the function $\Psi(\rho)$ is not
given by local scale-invariance. Rather, for $\rho\ll 1$ we expect that the
response of the two-time autocorrelation function $C(t,s)=C(s,t)$ should be 
symmetric and especially non-singular in the limit $t-s\to 0$ 
\cite{Pico04}. This suggests that $\Psi(\rho)\simeq\Psi_0\rho^{\lambda_C-d/2}$
if $\rho\leq\eps$ and $\Psi(\rho)$ is given by (\ref{gl:PsiConf}) only if
$\rho\geq\eps$ where  
$\eps$ sets the scale which separates the two regimes. The constant $\Psi_0$
is determined from the condition that $\Psi(\rho)$ is continuous at $\rho=\eps$.
A straightforward but slightly lengthy calculation gives
\BEA
\lefteqn{ \Phi(w) = B\left[ \left(\Gamma(d/2)\,
{_2F_1}(\lambda_C-d/2,d/2;1+2a+d/2-\lambda_C;-1/w)
-\gamma(d/2,Ew)\right)w^{-d/2}\right] }
\nonumber \\
& & + A \left[ \Gamma(\lambda_C-2a)\,
{_2F_1}(2\lambda_C-d-2a,\lambda_C-2a;\lambda_C+1-2a-d/2;-1/w)w^{2a-\lambda_C}
\right. 
\nonumber \\
& & ~~~ \left. -\gamma(\lambda_C-2a,Ew)w^{2a-\lambda_C}\right] 
+ A E^{-2a} w^{-\lambda_C}\gamma(\lambda_C,Ew) + 
     BE^{d/2-\lambda_C} w^{-\lambda_C}\gamma(\lambda_C,Ew) 
\nonumber \\
& & + A E^{1-2a}
\frac{2\lambda_C-d-2a}{\lambda_C+1-2a-d/2}w^{-\lambda_C}
\left[ (Ew)^{2a-1}\gamma(\lambda_C+1-2a,Ew) -\gamma(\lambda_C,Ew)\right]
\label{gl:PhiFinal} 
\EEA
where $\gamma(a,z)$ is an incomplete gamma-function, 
$E={\cal M}\eps$ and $A,B$ are constants related to $\psi_{0,1}$. 
Eqs.~(\ref{gl:CPhi},\ref{gl:PhiFinal}) together give the autocorrelation 
function $C(t,s)$. This is our main result. 

Some simple consistency checks are easy to perform. First, for free fields
$\lambda_C=d/2$ and eq.~(\ref{gl:appr}) is recovered for $A=0$. Second, we find
$\Phi(w)\sim w^{-\lambda_C}$ for $w\to\infty$ as expected and if $A\ne 0$,
we obtain the additional constraint $2a\leq 1$. 

\begin{figure}[t]
\centerline{\epsfxsize=3.5in\epsfbox
{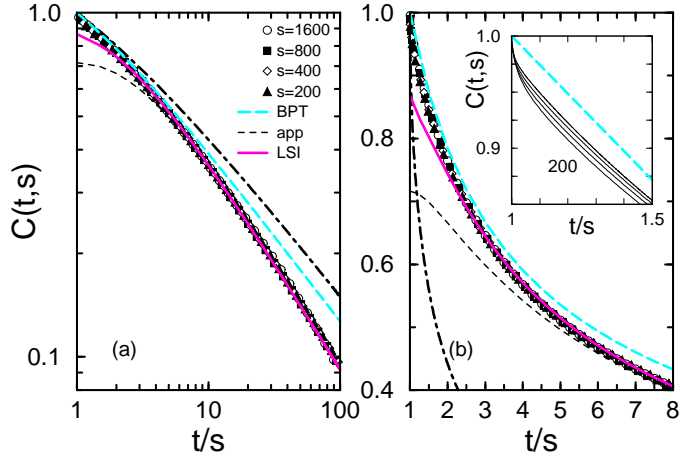}
}
\caption{Scaling of the autocorrelation function $C(t,s)$ 
of the $2D$ Glauber-Ising model at $T=0$. 
The curves are as follows: BPT corresponds to 
{\protect (\ref{gl:BPT})}, app to (\ref{gl:appr}) and 
LSI to (\ref{gl:CPhi},\ref{gl:PhiFinal}). Error bars are much smaller
than the symbol sizes. The dash-dotted line in a) gives a $2^{\rm nd}$-order 
perturbative correction of {\protect (\ref{gl:BPT})} \cite{Maze98} and the 
dash-dotted line in b) is the closed-form approximation of \cite{Liu91}. 
The inset shows $C(t,s)$ for $s=200, 400, 800$ and
$1600$ (from bottom to top), as well as the BPT line eq.~(\ref{gl:BPT}). 
\label{Bild1}}
\end{figure}

For a non-trivial test, we consider the phase-ordering kinetics of the 
$2D$ Glauber-Ising model, which we realize through a standard 
heat-bath rule, and lattices up to $800^2$. 
We consider quenches to temperatures $T=0$ and $T=1.5$, both
in the ordered phase. For $T=0$ ($T=1.5$) we went up to $y= t/s = 100$ ($y=60$),
with $s=1600$ being the longest waiting time studied, 
and averaged over typically
500 independent runs for the largest lattices. While the exponent $a=1/2$ is 
known \cite{Godr02,Henk02a}, we repeated the determination of $\lambda_C$
and find $\lambda_C=1.25(1)$, in agreement 
with earlier results \cite{Fish88,Brow97}. 
Next, we determined the amplitude $C_{\infty}$ from $C(t,s)\simeq C_{\infty}
(t/s)^{-\lambda_C/2}$ as $t/s\to\infty$ which produces a first constraint for
the fit of the constants $A,B,E$. A second constraint follows from the 
observation that at a special value $y_{\rm ref}$ the curves for different 
values of $s$ cross. We write $C(y_{\rm ref})=C_{\rm ref}$. 
The results are listed in
table~\ref{tab1}, together with the values of the parameters $A,B,E$ into
which we absorbed the normalization constant $a_0$. 
In figure~\ref{Bild1}, we finally compare our Monte Carlo data obtained for 
$T=0$ with several theoretical predictions. Clearly, the
waiting times considered are large enough to be inside the dynamical 
scaling-regime.  

Considering first a large range of values of $t/s$ (see figure~\ref{Bild1}a)
we observe that although the prediction (\ref{gl:BPT}) 
\cite{Bray91a,Bray92,Roja99} is quite close to the data for $t/s$ small 
(even so it lies systematically above the numerical data), there
is a strong deviation for $t/s\gtrsim 3$, see \cite{Brow97}. 
On the other hand, the simple
approximation (\ref{gl:appr}) works well for $t/s$ large but not surprisingly 
fails for $t\gtrsim s$ since the model at hand is not described by a free field.
Finally, local scale-invariance (LSI) as given by 
eqs.~(\ref{gl:CPhi},\ref{gl:PhiFinal}) and the parameters of table~\ref{tab1}
produces a nice overall agreement with the data, up to the smallish region
$t/s\lesssim 2$. That region is examined closer in figure~\ref{Bild1}b. 
We remark that for $t\simeq s$ dynamical scaling 
no longer holds true \cite{Zipp00}
(see the inset in figure~\ref{Bild1}b) and we cannot hope to be able to
find $C(t,s)$ from a dynamical symmetry argument. The
approximate analytical theories proposed in \cite{Maze98,Liu91} are only 
qualitatively correct which indicates that OJK-style approximations might 
not capture fully the quantitative aspects of phase-ordering. 

\begin{figure}[tb]
\centerline{\epsfxsize=3.5in\epsfbox
{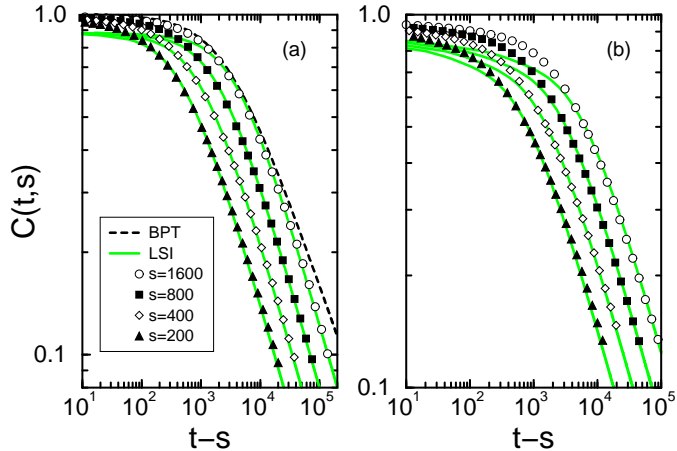}
}
\caption{Autocorrelator of the $2D$ Glauber-Ising model at temperatures 
(a) $T=0$ and (b) $T=1.5$. LSI is the prediction 
eqs.~(\ref{gl:CPhi},\ref{gl:PhiFinal}) and for $s=1600$ the curve BPT
eq.~(\ref{gl:BPT}) is also shown. 
\label{Bild2}}
\end{figure}

In figure~\ref{Bild2}, we present $C(t,s)$ in a more traditional way usually
preferred by experimentalists, for both $T=0$ and $T=1.5$, which makes the
simultaneous dependence of $C(t,s)$ on $t-s$ and on $s$ explicit. Again,
we find a nice agreement between LSI and the numerical data, but with
larger finite-time corrections to scaling for $T=1.5$ than for $T=0$. 
This finding is strong evidence that the extension of dynamical scaling to
Schr\"odinger-invariance and further to conformal invariance involving also
the `masses' as variables is indeed a true dynamical symmetry of phase-ordering
kinetics for all temperatures $T<T_c$. We recall that the LSI-prediction
(\ref{gl:CPhi},\ref{gl:PhiFinal}) depends on $\phi$ and $\wit{\phi}^2$ being
quasiprimary under Schr\"odinger/conformal transformations \cite{Henk02}. It
has turned out that the magnetic order-parameter of the Glauber-Ising model
is indeed quasiprimary. For the XY-model, however, the spin magnetization 
$\vec{S}(t,\vec{r})$ cannot be identified with a quasiprimary field but the
spin-wave approximation suggests that the phase variable $\phi(t,\vec{r})$
should take that r\^ole \cite{Pico04}. 

Summarizing, we have proposed to extend the usual dynamical scaling found
in phase-ordering kinetics to a {\em local} scale-invariance by (i) requiring
Galilei-invariance at $T=0$ and (ii) considering the dimensionful `masses'
of the order-parameter and response fields as further variables. This has led
us to postulate a new kind of time-dependent conformal invariance in  
phase-ordering kinetics. We have derived the explicit prediction 
eqs.~(\ref{gl:CPhi},\ref{gl:PhiFinal}) for the two-time autocorrelation 
function. This expression is in agreement with numerical results of the
$2D$ Glauber-Ising model and also agrees with 
several exactly solvable systems \cite{Pico04}. 

\acknowledgments
This work was supported by the Bayerisch-Franz\"osisches Hochschulzentrum
(BFHZ), by CINES Montpellier (projet pmn2095), and by NIC J\"{u}lich
(Projekt Her10).



\end{document}